\documentclass[fleqn,12pt,twoside]{article}
\usepackage[headings]{espcrc1}

\usepackage{graphicx}
\usepackage{epsfig,latexsym}

\newcommand{\be}{\begin{eqnarray}}
\newcommand{\ee}{\end{eqnarray}}

\readRCS
$Id: espcrc1.tex,v 1.2 2004/02/24 11:22:11 spepping Exp $
\usepackage{graphicx}
\usepackage[figuresright]{rotating}

\newcommand{\AmS}{{\protect\the\textfont2
  A\kern-.1667em\lower.5ex\hbox{M}\kern-.125emS}}

\title{Heavy Quark Correlators Above Deconfinement}

\author{\'Agnes M\'ocsy\address[MCSD]{Frankfurt
Institute for Advanced Studies \& Theoretical Physics Institute\\
J.W. Goethe-Universit\"at, Max von Laue Str 1, 60438 Frankfurt am
Main, Germany}%
       }

\runtitle{Heavy Quark Correlators Above Deconfinement}
\runauthor{\'A. M\'ocsy}

\begin{document}
\maketitle

\begin{abstract}
We compare the quarkonia correlators from potential models and
from lattice QCD.
\end{abstract}

\section*{Why Interested in Heavy Quark Correlators?}

There has been lots of interest in studying the modifications of
the heavy quark bound states, ever since in 1986 Matsui and Satz
predicted \cite{Matsui:1986dk} the suppression of the $J/\psi$
peak in the dilepton spectra as signal for deconfinement. The idea
was that color screening by the light quarks in the deconfined
phase would prevent the binding of a heavy quark and antiquark.
Potential model calculations predicted that higher excitations
disappear earlier \cite{Karsch:1987pv}, and found that the
$J/\psi$ melts at about $1.1T_c$ \cite{Digal:2001ue}.

The recently available evaluation on the lattice of quarkonia
correlators and spectral functions
\cite{Umeda:2002vr,Datta:2003ww,Petrov:2005sd} presented a
surprise: The 1S $J/\psi$ and $\eta_c$, survive at least up to
$1.5T_c$, with no significant temperature-dependence of their
masses \cite{Umeda:2002vr,{Datta:2003ww}}, the 1S $\eta_b$ is not
modified until about $2T_c$ \cite{Petrov:2005sd}, the 1P
$\chi_c^0$ and $\chi_c^1$ are dissolved at $1.1T_c~$
\cite{Datta:2003ww}, and the $\chi_b^0$ shows drastic changes
already at $1.15T_c$ \cite{Petrov:2005sd}.

While the lattice results do not contradict the idea of sequential
dissolution, they are not in agreement with previous potential
model calculations. It is questionable though, whether medium
modifications of quarkonia can be understood in terms of a
temperature dependent potential. Also, even though the extraction
of spectral functions from the lattice is very promising, the
currently available results are not yet fully reliable. The
correlators, however, are accurately determined on the lattice.

Here we present some of our results obtained when investigating
heavy quark correlators in potential models with a screened
Cornell potential, as well as a potential fitted to the internal
energy obtained on the lattice \cite{Mocsy:2004bv,new}, and we
compare them to the correlators obtained on the lattice.

\section*{Potential Models}

Potential models intend to understand the modifications at finite
temperature of the quarkonia properties using some screened
potential to mediate the interaction between the heavy quark and
antiquark \cite{yellow}. A usual screened potential is a finite
temperature extension of the zero temperature Cornell potential
\cite{Eichten:1979ms}. We used two, formally very different
potentials: One is a phenomenologically motivated screened
potential \cite{Karsch:1987pv}
\be
V(r,T)=-\frac{\alpha}{r}e^{-\mu(T) r}+\frac{\sigma}{\mu(T)}
(1-e^{-\mu(T) r}) \, , \label{pot}\ee with screening mass
 $\mu=[0.24+0.31 \cdot(T/T_c-1)]~$GeV,
$T_c=0.270~$GeV, coupling constant $\alpha=0.471$, and string
tension $\sigma=0.192$GeV$^2$. The other is a recently popular
\cite{Shuryak:2003ty}, yet still questionable identification as
potential of the internal energy of a static $q\bar{q}$ pair
determined on the lattice. We fit the lattice data on the internal
energy from \cite{Kaczmarek:2003dp} by the Ansatz
\be
V(r,T) &=& -\frac{\alpha}{r}e^{-\mu(T) r^2} + \sigma(T) r
e^{-\mu(T) r^2} + C(T)\left(1-e^{-\mu(T) r^2}\right)\,
.\label{latpot}\ee Solving the Schr\"odinger equation with these
potentials provides the binding energies, the radii, and the wave
function of the different bound states. Importantly, the results
turn out to be not sensitive to the details of the potential. The
masses obtained are in accordance with the lattice, and do not
change substantially with temperature, except for the scalar
charmonium $\chi_c^0$, whose properties are modified already at
$1.1T_c$ \cite{Mocsy:2004bv}. All the wave functions at the origin
show a strong drop with increasing temperature
\cite{Mocsy:2004bv}. The properties of the pseudoscalar $\eta_c$
and the vector $J/\psi$ states are identical, since we neglect
effects that would arise from the hyperfine splitting.

It is common for all screened potentials that these are finite at
infinite separation distances between the quark and antiquark.
This determines a threshold for the continuum,
$s_0(T)=2m+\mbox{lim}_{r\rightarrow\infty} V(r,T)$, above which
the quark/antiquark can propagate freely. This threshold decreases
with increasing temperature.

Assuming thus, that the a heavy quark and antiquark in the
deconfined phase interact via a screened potential $V(r,T)$, and
propagate freely above the threshold $s_0(T)$, we motivate the
following form for the spectral function \cite{Mocsy:2004bv,new}
\be
\sigma(\omega,T) = \sum_i 2 M_i(T) F_i(T)^2 \delta\left(\omega^2
-M_i^2\right) + \frac{3}{8\pi^2} \omega^2
\theta\left(\omega-s_0(T)\right) +
\chi_s\left(1-3\frac{T}{M}\right)\omega\delta(\omega) \, .
\label{spft} \ee $M_i(T)$ and $F_i(T)$ are the bound state masses
and amplitudes. The last term, present only in the vector channel,
is due to charge fluctuations and diffusion, with $\chi_s(T)$ the
charge susceptibility \cite{new}. For the continuum here we chose
a sharp threshold, which needs not be the case \cite{new}. The
correlator is now obtained from its spectral representation
\be
G(\tau,T)=\int d\omega \sigma(\omega,T)
\frac{\cosh{\left[\omega\left(\tau-\frac{1}{2T} \right)\right]}}
{\sinh{\left[\frac{\omega}{2T}\right]}}\, . \label{corr}\ee To
make direct comparison with the lattice results, we determine the
ratio $G/G_{recon}$, where $G_{recon}$ is the correlator
(\ref{corr}) evaluated with $\sigma(\omega,T=0)$.

\section*{Results and Conclusions}

\begin{figure}[htbp]
\begin{minipage}[h]{5cm}
\epsfig{file=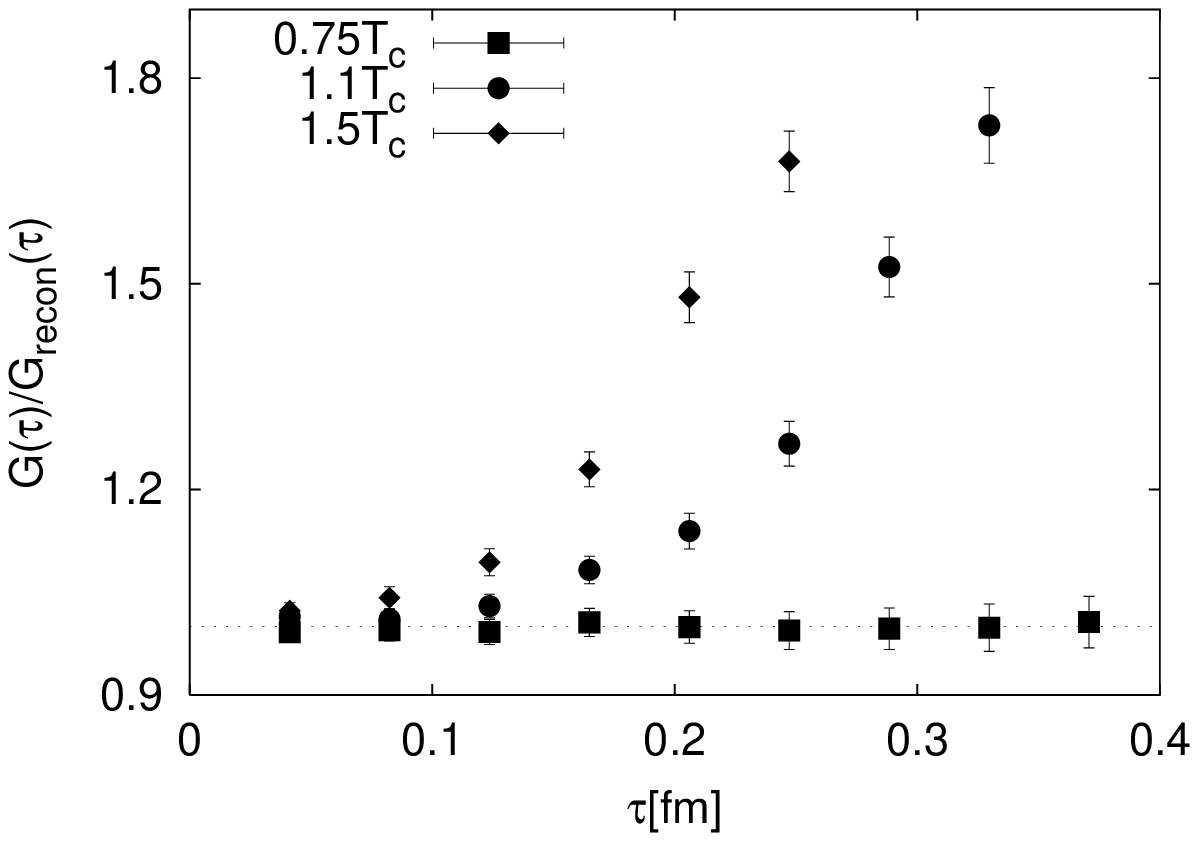,height=46mm}
\end{minipage}
\hspace*{2cm}
\begin{minipage}[h]{5cm}
\epsfig{file=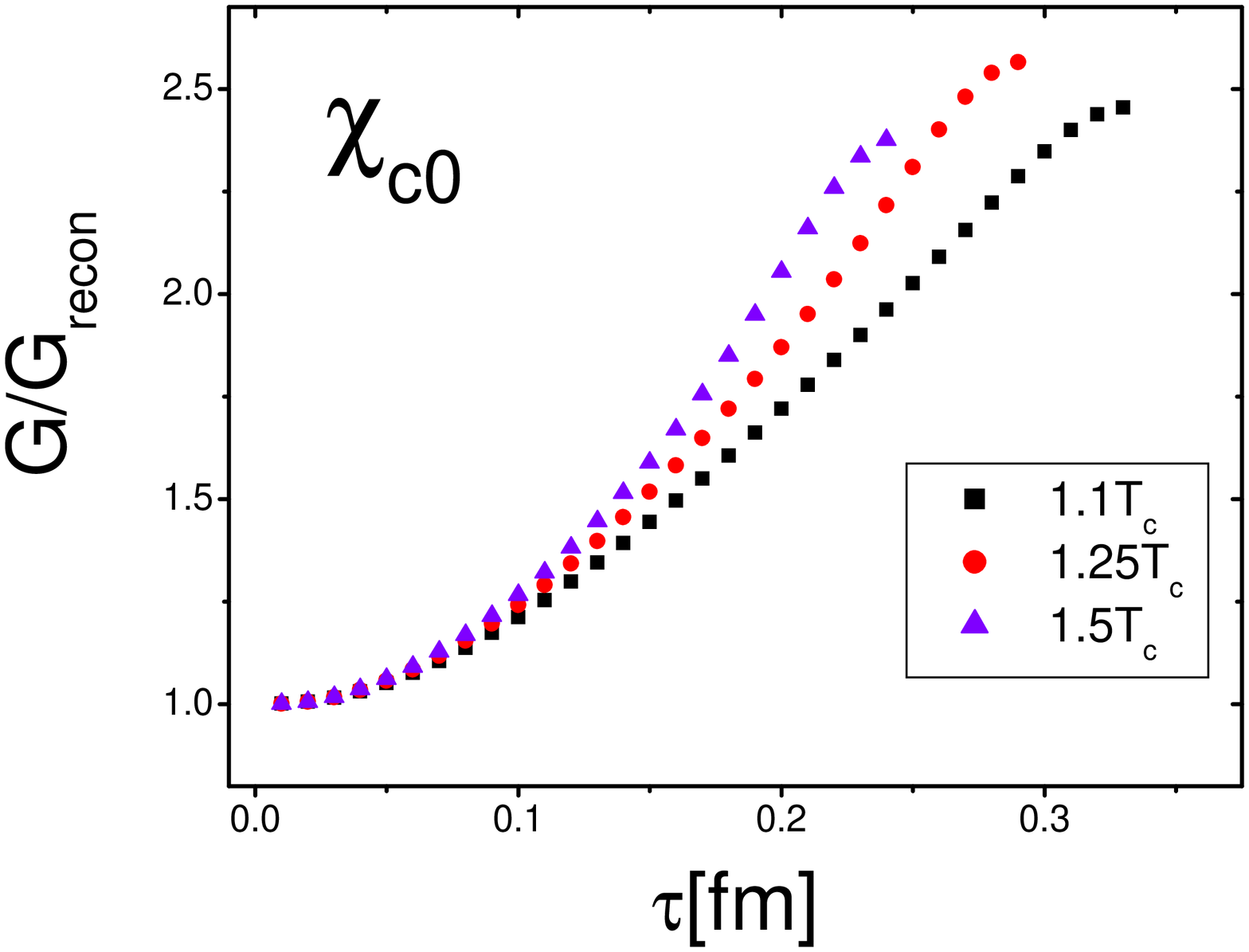,height=44mm}
\end{minipage}\\
\begin{minipage}[h]{5cm}
\epsfig{file=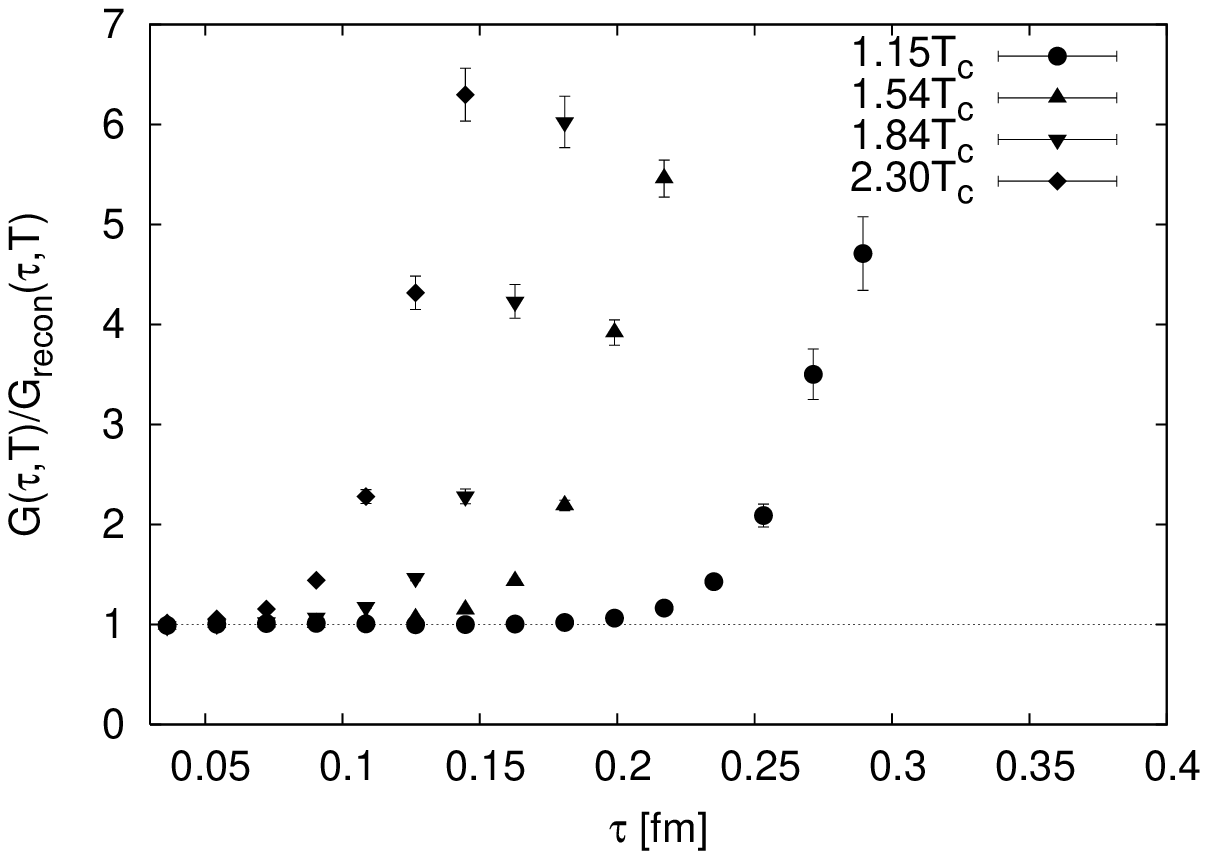,height=44mm}
\end{minipage}
\hspace*{2cm}
\begin{minipage}[h]{5cm}
\epsfig{file=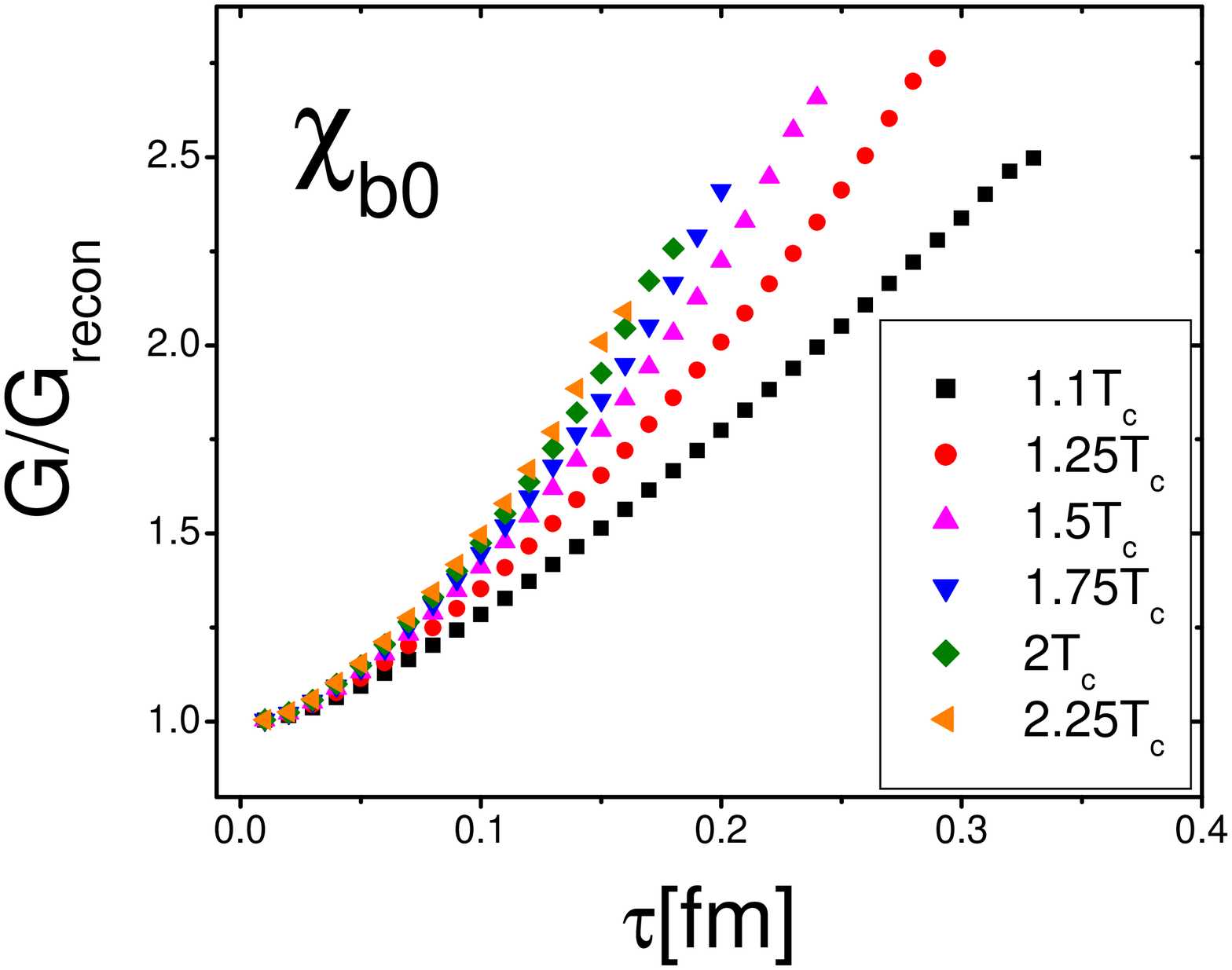,height=44mm}
\end{minipage}\\
\begin{minipage}[h]{5cm}
\epsfig{file=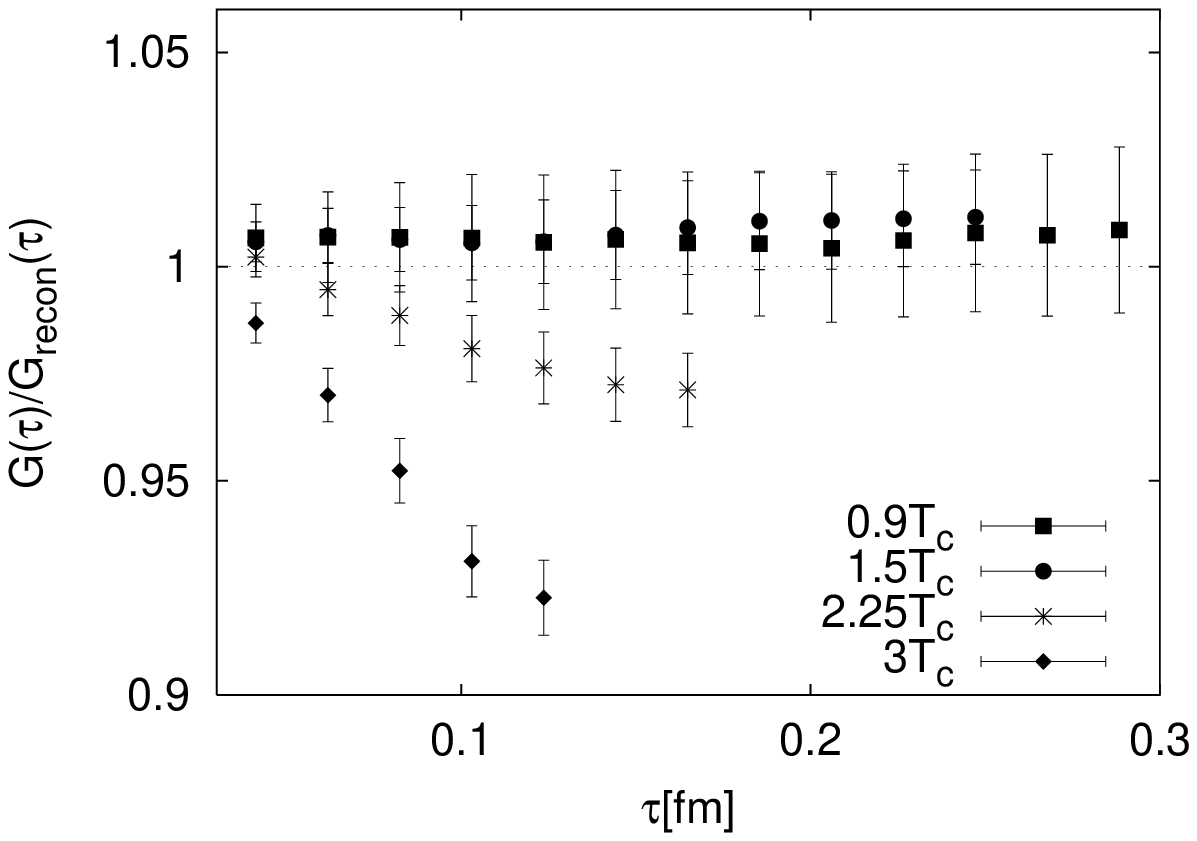,height=46mm}
\end{minipage}
\hspace*{2cm}
\begin{minipage}[h]{5cm}
\epsfig{file=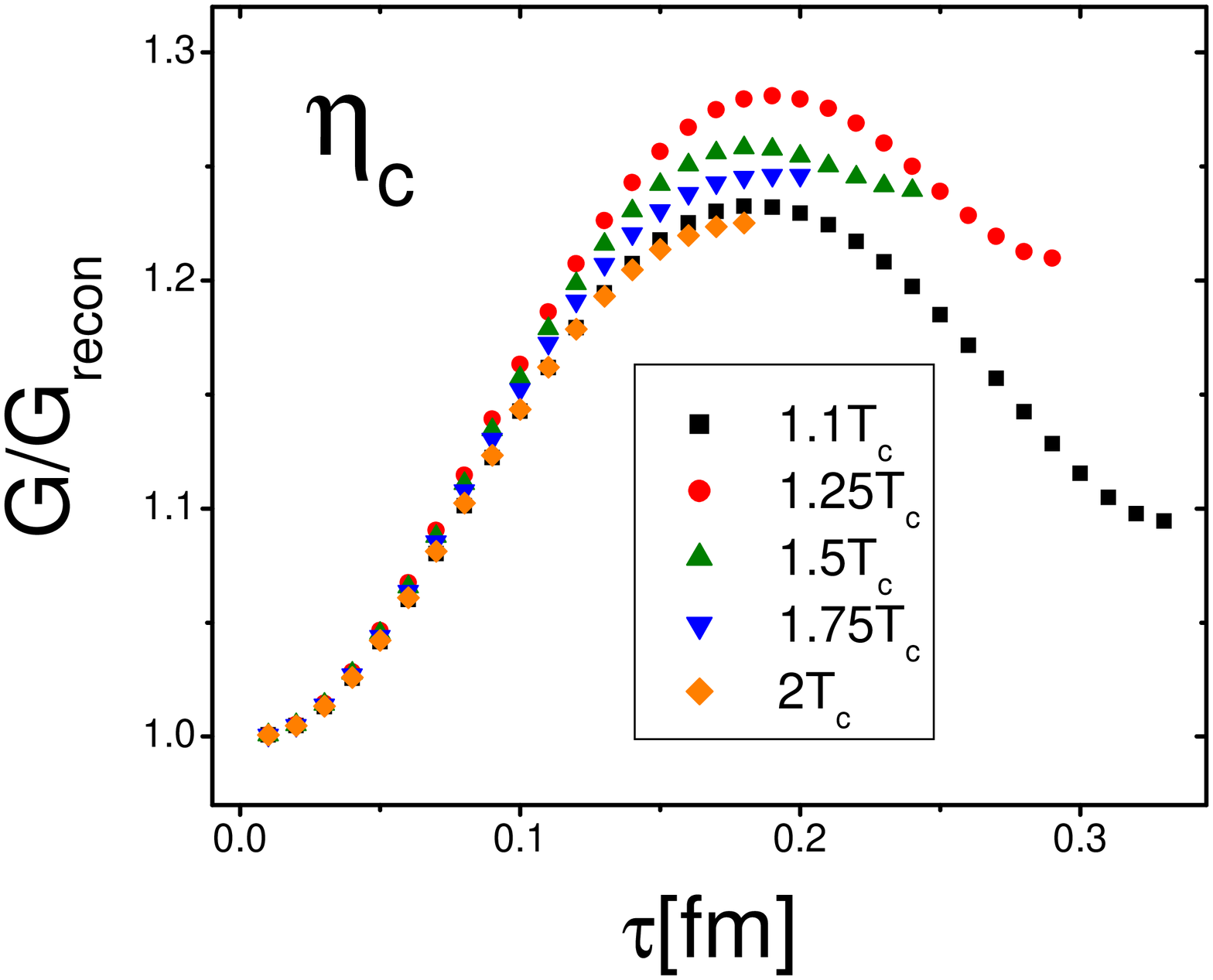,height=44mm}
\end{minipage}\\
\begin{minipage}[h]{5cm}
\epsfig{file=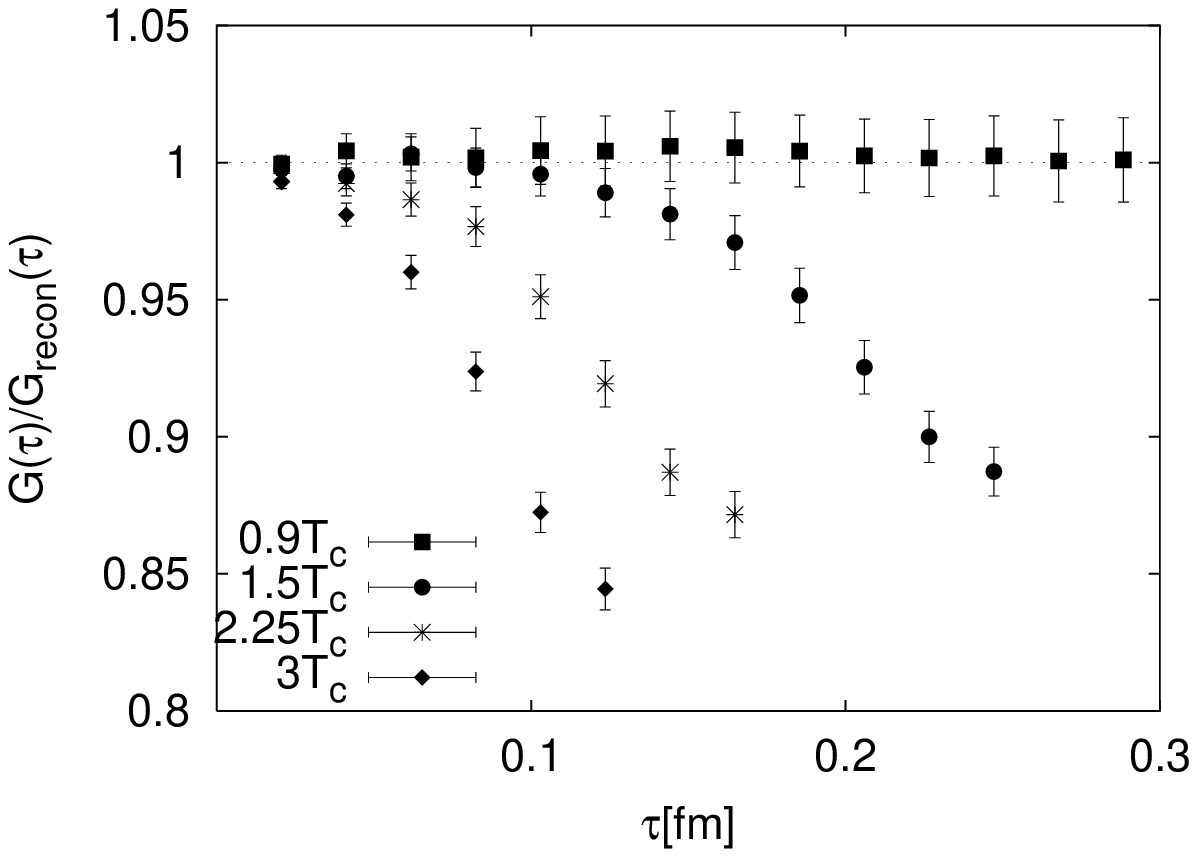,height=46mm}
\end{minipage}
\hspace*{2cm}
\begin{minipage}[h]{5cm}
\epsfig{file=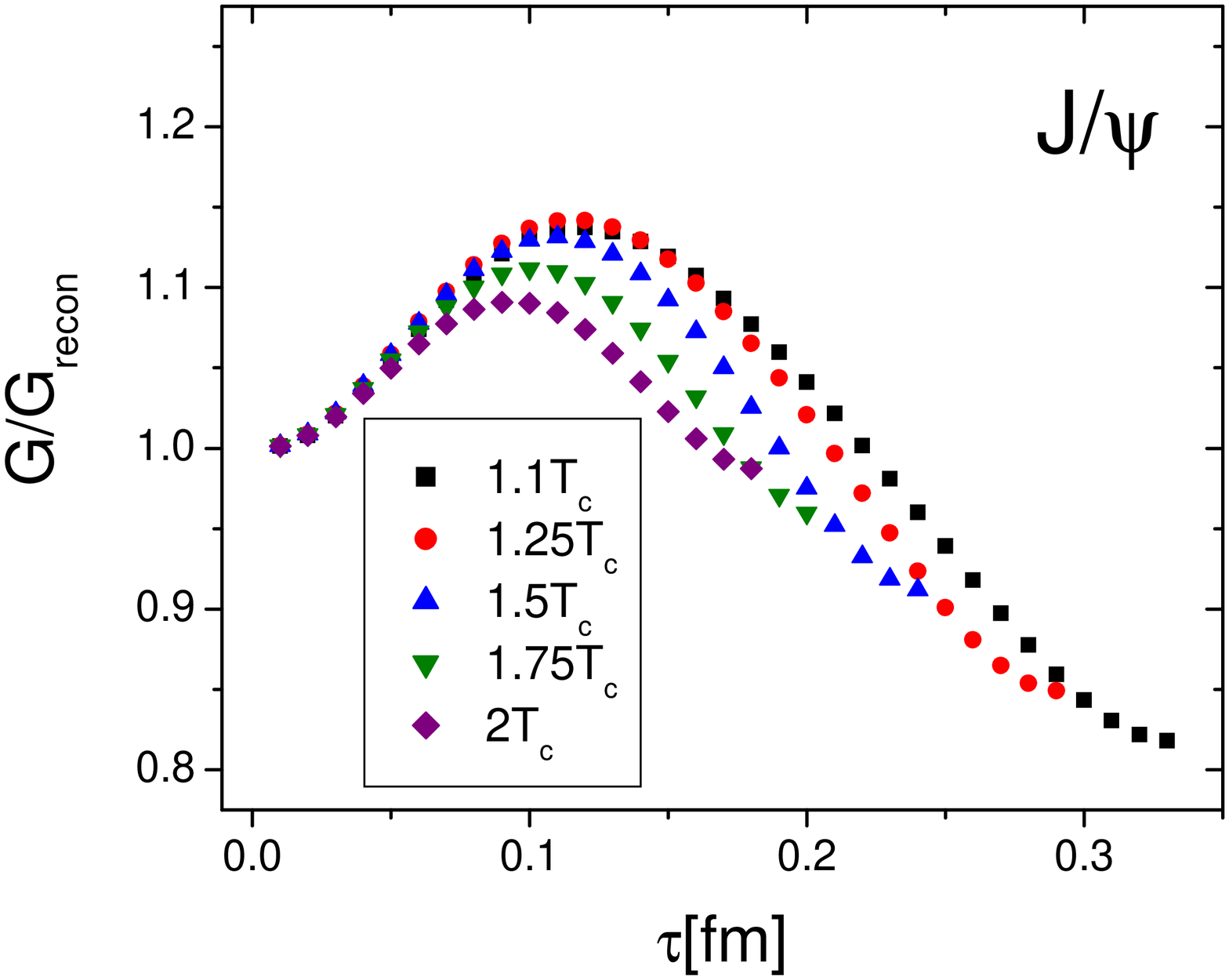,height=44mm}
\end{minipage}
\caption{Top to bottom: Ratio of the $\chi_c^0$, $\chi_b^0$,
$\eta_c$ and $J/\psi$ correlator to the reconstructed correlator
calculated on the lattice (left panels from
\cite{Datta:2003ww,{Petrov:2005sd}}) and in our model (right
panels). } \label{fig}
\end{figure}

Some results for the correlators are summarized in Figure
\ref{fig} as obtained on the lattice (left panels) and in our
potential model calculations (right panels). For the scalar
charmonium (top panel) an increase of the correlator has been
detected right above $T_c$. The increase is enhanced with
temperature, despite the fact that the contribution from the
$\chi_c^0$ state is negligible. This enhancement is due to the
thermal shift of the continuum threshold.
The scalar bottomonium correlator also shows an increase (top 2nd
panel) right above $T_c$, even though the $\chi_b^0$ survives
until much higher temperatures than the $\chi_c^0$. Thus the
continuum seems to be the dominant contribution to the scalar
correlator.

Since in our calculations the pseudoscalar and vector channel
correspond to the same 1S state, we would not expect the $\eta_c$
and the $J/\psi$ correlator to behave differently. The lattice
correlator shows no change up until about $3T_c$ for the $\eta_c$
(2nd bottom left panel), yet a significant decrease for the
$J/\psi$ already at $1.5T_c$ at distances $>0.15~$fm
(c.f.~eq.~(\ref{spft}) and bottom left panel). We explain this
with the effects of diffusion and charge fluctuations that make
the $J/\psi$ correlator smaller. Our model calculations show this
effect too (bottom two right panels). However, for both the
$\eta_c$ and the $J/\psi$, the potential model calculations do not
reproduce the behavior of the correlator as obtained from the
lattice. The identified extra feature is the significant
contribution to the correlator from the continuum due to threshold
reduction, and manifest in the increase of the correlators at
short distances.

The features identified are independent of the form of the
potential. The temperature dependence of the correlators show much
richer structure than the one seen on the lattice. We are thus led
to the question whether some physics is not identified in the
lattice correlators due to lattice artifacts, or the physics in
the deconfined phase is more subtle than allowing to describe
medium effects on heavy quark bound states by potential models.
Further investigations of these questions are in progress.

\section*{Acknowledgements}

This presentation is based on work done in collaboration with
P\'eter Petreczky. I thank the Humboldt Foundation for financial
support, and S.~Datta for the lattice figures.


\begin{thebibliography}{50}

\bibitem{Matsui:1986dk}
T.~Matsui and H.~Satz,
Phys.\ Lett.\ B {\bf 178}, 416 (1986).

\bibitem{Karsch:1987pv}
F.~Karsch, M.~T.~Mehr and H.~Satz,
Z.\ Phys.\ C {\bf 37}, 617 (1988).

\bibitem{Digal:2001ue}
S.~Digal, P.~Petreczky and H.~Satz,
Phys.\ Rev.\ D {\bf 64}, 094015 (2001) [arXiv:hep-ph/0106017].

\bibitem{Umeda:2002vr}
T.~Umeda, K.~Nomura and H.~Matsufuru,
Eur.\ Phys.\ J.\ C {\bf 39S1}, 9 (2005) [arXiv:hep-lat/0211003];
M.~Asakawa and T.~Hatsuda,
Phys.\ Rev.\ Lett.\  {\bf 92}, 012001 (2004)
[arXiv:hep-lat/0308034].

\bibitem{Datta:2003ww}
S.~Datta, F.~Karsch, P.~Petreczky and I.~Wetzorke,
Phys.\ Rev.\ D {\bf 69}, 094507 (2004) [arXiv:hep-lat/0312037].

\bibitem{Petrov:2005sd}
K.~Petrov,
arXiv:hep-lat/0503002;
A.~Jakov\'ac, P.~Petreczky, K.~Petrov and A.~Velytsky, work in
progress.

\bibitem{Mocsy:2004bv}
\'A.~M\'ocsy and P.~Petreczky,
Eur.\ Phys.\ J.\ C {\bf 43}, 77 (2005) [arXiv:hep-ph/0411262].

\bibitem{new}
\'A.~M\'ocsy and P.~Petreczky, in preparation.

\bibitem{yellow}
N. Brambilla et al, {\em Quarkonium Physics}, CERN Yellow Report,
hep-ph/0412158

\bibitem{Eichten:1979ms}
E.~Eichten, K.~Gottfried, T.~Kinoshita, K.~D.~Lane and T.~M.~Yan,
Phys.\ Rev.\ D {\bf 21}, 203 (1980).

\bibitem{Shuryak:2003ty}
E.~V.~Shuryak and I.~Zahed,
Phys.\ Rev.\ C {\bf 70}, 021901 (2004) [arXiv:hep-ph/0307267];
Phys.\ Rev.\ D {\bf 70}, 054507 (2004) [arXiv:hep-ph/0403127];
C.~Y.~Wong,
arXiv:hep-ph/0408020;
W.~M.~Alberico, A.~Beraudo, A.~De Pace and A.~Molinari,
arXiv:hep-ph/0507084.

\bibitem{Kaczmarek:2003dp}
O.~Kaczmarek, F.~Karsch, P.~Petreczky and F.~Zantow,
Nucl.\ Phys.\ Proc.\ Suppl.\  {\bf 129}, 560 (2004)
[arXiv:hep-lat/0309121].


\end{thebibliography}
\end{document}